\documentclass[a4paper,fleqn]{cas-dc}

\usepackage[numbers]{natbib}

\usepackage{xcolor}
\usepackage{graphicx}
\usepackage{float}
\usepackage{amsmath}
\usepackage{amssymb}

\def\tsc#1{\csdef{#1}{\textsc{\lowercase{#1}}\xspace}}
\tsc{WGM}
\tsc{QE}
\tsc{EP}
\tsc{PMS}
\tsc{BEC}
\tsc{DE}

\begin{document}
\let\WriteBookmarks\relax
\def\floatpagepagefraction{1}
\def\textpagefraction{.001}
\shorttitle{LINEAR: Learning Implicit Neural Representation With Explicit Physical Priors for Accelerated Quantitative \texorpdfstring{$\text T_{1\rho}$}{T1rho} Mapping}
\shortauthors{Y.Liu, J.Xie, and Z.Cui \textit{et~al.}}

\title [mode = title]{LINEAR: Learning Implicit Neural Representation With Explicit Physical Priors for Accelerated Quantitative \texorpdfstring{$\text T_{1\rho}$}{T1rho} Mapping}


\author[1]{\textcolor{black}{Yuanyuan Liu}}
\fnmark[1] 
\author[2]{\textcolor{black}{Jinwen Xie}}
\fnmark[1]
\fntext[fn1]{Yuanyuan Liu and Jinwen Xie contributed equally to this work.}
\author[3]{\textcolor{black}{Zhuo-Xu Cui}}
\author[3]{\textcolor{black}{Qingyong Zhu}}
\author[1]{\textcolor{black}{Jing Cheng}}
\author[3]{\textcolor{black}{Dong Liang}}[orcid=0000-0003-0131-2519]
\ead{dong.liang@siat.ac.cn} 
\author[1]{\textcolor{black}{Yanjie Zhu \corref{cor1}}}[orcid=0000-0002-9131-689X]
\cortext[cor1]{Correspongding author: Dong Liang; Yanjie Zhu}
\ead{yj.zhu@siat.ac.cn}


\affiliation[1]{organization={Paul C.Lauterbur Research Center for Biomedical Imaging, Shenzhen Institute of Advanced Technology, Chinese Academy of Sciences}, city={Shenzhen}, country={China}}
\affiliation[2]{organization={Software Engineering Institute, East China Normal University}, city={Shanghai}, country={China}}
\affiliation[3]{organization={Research Center for Medical AI, Shenzhen Institute of Advanced Technology, Chinese Academy of Sciences}, city={Shenzhen}, country={China}}


\begin{abstract}
Quantitative $\text T_{1\rho}$  mapping has shown promise in clinical and research studies. However, it suffers from  long scan times. Deep learning-based techniques have been successfully applied in accelerated quantitative MR parameter mapping. However, most methods  require  fully-sampled training dataset, which is impractical in the clinic. In this study, a novel subject-specific unsupervised method based on the implicit neural representation is proposed to reconstruct $\text T_{1\rho}$-weighted images from highly undersampled $k$-space data, which only takes spatiotemporal coordinates as the input. Specifically, the proposed method learned a implicit neural representation of the MR images driven by two explicit priors from the physical model of $\text T_{1\rho}$ mapping, including the signal relaxation prior and self-consistency  of $k$-t space data prior. The proposed method was verified using both retrospective and prospective undersampled $k$-space data. Experiment results demonstrate that LINEAR achieves a high acceleration factor up to 14, and outperforms the state-of-the-art methods in terms of suppressing artifacts and achieving the lowest error.
\end{abstract}



\begin{keywords}
Implicit neural representation\sep Relaxation and self-consistency priors\sep Parameter mapping
\end{keywords}

\maketitle

\section{Introduction}
Quantitative magnetic resonance (QMR) parameter mapping, including $\text{T}_1$, $\text{T}_2$, and $\text T_{1\rho}$ mapping, has shown great potential in diagnosis over various clinical applications~\cite{ma2013MRF,ma2021MTP,lescher2015quantitative,wang2015t1rho}. $\text T_{1\rho}$ mapping is an emerging technique that describes spin-lattice relaxation in the rotating frame~\cite{wang2015t1rho}. It has been utilized in studying cartilage pathology in osteoarthritis, early diagnosis of degenerative intervertebral disc disease, and neurological disorders including Alzheimer's disease, multiple sclerosis, stroke, and Parkinson's disease~\cite{wang2015t1rho, haris2011t1rho,mangia2014magnetization,jokivarsi2010estimation,nestrasil2010t}. However, $\text T_{1\rho}$ mapping requires the acquisition of multiple $\text T_{1\rho}$-weighted images with varying spin-lock times (TSL), leading to a long scan time, which greatly hinders its widespread clinical use. Therefore, accelerated quantitative $\text T_{1\rho}$ mapping is highly desired.
\par Traditional accelerated acquisition methods like parallel imaging (PI) and compressed sensing (CS) have been successfully used in accelerated quantitative MR parameter mapping through undersampling $k$-space data. The PI techniques which use sensitivity information to remove aliasing artifacts resulting from undersampling has been employed in clinical practice~\cite{pruessmann1999sense,griswold2002generalized}. In CS, original signals can be recovered from highly undersampled $k$-space data by exploiting transform domain sparsity~\cite{lustig2007sparse, block2007undersampled, lingala2013blind, doneva2010compressed} or low-rank properties of the image and $k$-space~\cite{ otazo2015low, jin2016general, shin2014calibrationless, haldar2013low, haldar2020linear, lee2016acceleration}. However, acceleration rate of PI and CS is limited by the reduction of signal-to-noise ratio (SNR) and degradation of image quality, making it challenging to meet the requirements of quantitative $\text T_{1\rho}$ mapping.

Deep learning (DL) has shown promising results in MR reconstruction from highly undersampled $k$-space data. Unlike PI and CS-based methods, DL-based approaches utilize image priors encoded in trained network weights, learned from training data~\cite{zhu2023physics}. These methods can be categorized into supervised and unsupervised learning. Supervised learning methods train networks using paired data to map undersampled $k$-space or aliased images to fully sampled $k$-space or artifact-free images~\cite{chung2022score,knoll2020deep}. Unsupervised learning methods, on the other hand, either learn data probability distribution using networks (e.g., GANs or diffusion models) or train networks using holdout masking operation on the acquired measurement data data~\cite{yaman2020self,akccakaya2022unsupervised, cui2022self}. However, most DL-based methods require large-scale training datasets with paired fully sampled $k$-space data~\cite{liang2020deep}, which may not be feasible in practical imaging scenarios.

Recently, a method known as implicit neural representation (INR) has emerged as a promising solution for image reconstruction without ground-truth high-quality images~\cite{mildenhall2021nerf,xu2023nesvor,sitzmann2020implicit}. Unlike traditional explicit representations that use discrete elements such as pixels (for 2D images) or voxels (for 3D volumes), INR parameterizes a continuous representation of the signal as an implicit function learned by a neural network, typically a multilayer perceptron (MLP).

This function maps spatial coordinates to the signal. INR offers advantages such as imposing implicit constraints on network output~\cite{linr2023}, including local continuity and global consistency. These properties align with high-quality images where neighboring pixels exhibit smooth changes. Consequently, the output adhering to these constraints closely resembles high-quality images. Additionally, prior knowledge, such as sparsity and low rank, can be integrated through appropriate loss functions.

Based on INR, several reconstruction methods have been developed to accelerate MR imaging. One straightforward way is using INR to represent the $k$-space data. Huang et al.~\cite{{shen2022nerp}} introduced neural implicit $k$-space (NIK) to interpolate the missing $k$-space data for non-cartesian cardiac MR imaging. However, the training of $k$-space INR faces significant challenges due to the large range and imbalanced distribution of $k$-space values. Another way is using INR to represent the image to be reconstructed. For example, Shen et. al~\cite{shen2022nerp} proposed an implicit neural representation learning with prior embedding (NeRP) to reconstruct images from undersampled $k$-space data with radial trajectories. However, NeRP requires a high-quality image from the same patient as the prior image to initialize INR, which limits its applicable clinical scenarios. Feng et al.~\cite{feng2023imjense} applied INR for joint coil sensitivity and image estimation (IMJENSE) in parallel imaging, achieving a net acceleration of up to 5. Shao et al.~\cite{shao20243d} used INR to accelerate 4D MR imaging by interpolating motion states during the scan. Nevertheless, only the sparsity constraint, i.e. total variation (TV) on the image is utilized in these methods, which limits their performance. In the context of QMR imaging, additional prior knowledge, such as the signal relaxation prior or image structural similarity prior, can be employed to further enhance the reconstruction performance. 

Therefore, we developed LINEAR, a subject-specific unsupervised reconstruction method for accelerated MR $\text T_{1\rho}$ mapping. 
This method learns an implicit neural representation of the $\text T_{1\rho} $-weighted images which jointly explores the correlations of $\text T_{1\rho}$-weighted images and multi-channel $k$-space data. 

The major contributions are:
\par (1) A new subject-specific unsupervised framework incorporating INR and physical prior is proposed to reconstruct $\text T_{1\rho} $-weighted images from highly undersampled $k$-space data, which is the first application of INR in MR $\text T_{1\rho} $ mapping. 

\par (2) We introduce a specific loss function for QMR which jointly explores the correlations of $\text T_{1\rho}$-weighted images and multi-channel $k$-space data across all TSLs. The proposed LINEAR method characterizes the signal relaxation prior and self-consistency of $k$-t space prior, significantly improving the reconstruction performance.

\par (3) The LINEAR method enables efficient $\text T_{1\rho}$ map estimation with reduced $k$-space samples, resulting in a reduction of scan time up to 14-fold. LINEAR outperforms the state-of-the-art unsupervised DL methods and traditional reconstruction methods, establishing its superiority in image reconstruction.

The paper is structured as follows. Section 2 reviews related works. Section 3 describes the proposed method. Implementation details are provided in Section 4. Section 5 presents the reconstruction results across various imaging scenarios and compares them with other state-of-the-art methods, analyzing the effects of the priors on reconstruction. Section 6 offers discussions, and Section 7 concludes the paper.

\begin{figure*}[!ht]
\centering{\includegraphics[width=2\columnwidth]{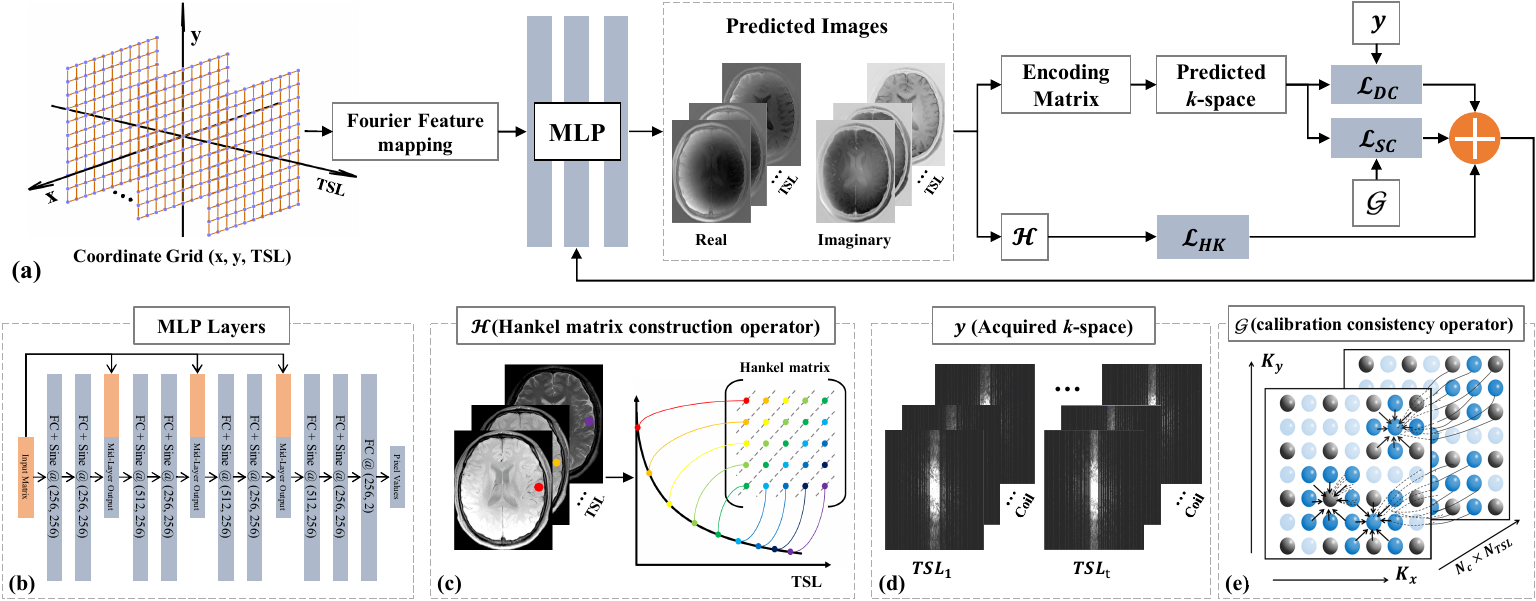}}
\caption{Flow diagram of the proposed LINEAR method, architecture of MLP, illustration of Hankel matrix construction process, undersampled $k$-space data for all TSLs and self-consistency of $k$-t space. (a) Flow diagram of LINEAR (b) detailed architecture of MLP (c) illustration of Hankel matrix construction process (d) an example of the undersampled $k$-space data at different TSLs (e) illustration of self-consistency of multi-channel $k$-space data across all TSLs.}
\label{fig1}
\end{figure*}

\section{Related work}
\subsection{Parallel imaging}
In parallel imaging, since $k$-space data are acquired with multiple receiver coils, the undersampled data can be recovered using knowledge of of coil sensitivity profile or autocalibration signals and the convolution kernel. According to the reconstruction implementation domain, the PI reconstruction methods can be categorized into two main types: image domain methods, such as sensitivity encoding (SENSE)~\cite{pruessmann1999sense} and $k$-space domain methods, such as generalized partially parallel acquisitions (GRAPPA)~\cite{griswold2002generalized}. SENSE utilizes prior knowledge of coil sensitivity profiles to separate folded pixels and reconstruct the image, while GRAPPA expresses a point in $k$-space as a linear combination of multi-coil signals and neighboring points. Advanced methods like iterative self-consistent parallel imaging (SPIRiT)~\cite{lustig2010spirit} and eigenvalue approach to autocalibrating PI (ESPIRiT)~\cite{uecker2014espirit} ensure self-consistency in $k$-space by leveraging correlated information from multiple coils and neighboring points. 

\subsection{Low rank property in QMR}
In QMR, image series are acquired with different pulse sequence parameters, and the signal variation of each image pixel across the dataset reflects the MR parameter to be estimated. Data redundancy exists along the temporal dimension in these images. This redundancy can be analyzed by forming an image matrix, or high-order tensor with low rank property~\cite{zhu2018scope, liu2023accelerating,zhang2015accelerating}. Zhao et al.~\cite{zhao2015accelerated} enforced low-rank structure of $\text T_2$-weighted images by constructing a Casorati matrix, where each column consists of the image pixels from each $\text T_2$-weighted image. Zhang et al.~\cite{zhang2015accelerating} restricted the Casorati matrix to a local image region and proposed a locally low rank method for better performance. Zhu et al.~\cite{zhu2018scope} proposed a signal compensation strategy to futher enhance the low rank of the image matrix with a rank value of 1. Bustin et al.~\cite{bustin2019high} proposed a high-dimensionality undersampled patch-based reconstruction method which exploited the data redundancy by constructing high-order low rank tensors. In the above approaches, the image reconstruction is regarded as a low-rank matrix or tensor completion problem. Moreover, additional constraints such as sparsity can be combined to enable reconstruction of images from highly undersampled data~\cite{,peng2016accelerated,zhou2022accelerating}.

\subsection{Implicit neural representation (INR) of MR images}
In INR, the $\text T_{1\rho}$-weighted images can be modeled as a continuous function parameterized by a neural network $f_{\theta}$ which takes the spatial coordinates as input and is implemented by an MLP. The neural network can be defined as
\begin{equation}
    \begin{gathered}
        f_{\theta}:\boldsymbol v \rightarrow \boldsymbol{I} \quad \text {with } \boldsymbol v \in \mathbb{R}^n, \boldsymbol I \in \mathbb{C},
    \end{gathered}
\end{equation}
where $\boldsymbol v = [v_{x}^{h},v_{y}^{w},v_{TSL}^k]_{h = 1,2,\cdots,N_x,w = 1,2,\cdots,N_y, k = 1,2,\cdots,N_{TSL}}$ is the coordinate space, $N_{x}$ and $N_{y}$ denote the number of frequency and phase encoding lines, respectively, $N_{TSL}$ is the number of TSLs, $\boldsymbol I$ is the corresponding complex signal value space, $n$ denotes the number of dimensions for coordinates (i.e. $n = 3$ in 2D imaging). Previous studies have indicated that deep networks operating directly on input coordinates perform inadequately in representing high-frequency variations. To address this issue, input coordinates can be encoded into a higher-dimensional space before being fed into the network using Fourier feature mapping to enhance the network's ability to learn high-frequency information~\cite{tancik2020fourier}. The encoded coordinates are defined as follows:
\begin{equation}
\begin{gathered}
    \gamma(\boldsymbol v)=[\cos (2 \pi \boldsymbol{B v}), \sin (2 \pi\boldsymbol{B v})]^{\mathrm{T}}
\end{gathered}
\label{fourier embed}
\end{equation}
where $\boldsymbol B \in \mathbb {R}^{n\times n_{e}}$ represents the coefficients for Fourier feature transformation, and $n_{e}$ is the dimension of Fourier features. The entries of $\boldsymbol{B}$ are independently sampled from the normal Gaussian distribution $\mathcal{N}(0, \sigma^2)$, where $\sigma$ is a hyperparameter characterizing the standard deviation~\cite{tancik2020fourier}. Additionally, using periodic activation functions, such as the sine activation function, enhances the MLP's ability to model intricate details of a signal and its derivatives~\cite{sitzmann2020implicit}.

\section{Proposed Method}
\subsection{Forward Model}
Let $ \boldsymbol{x }\in \mathbb{C}^{N_{v} \times N_{\textit {TSL}}}$ be the $\text T_{1\rho}$-weighted images to be reconstructed, where $N_v$ is the total number of pixels in the image ($N_v= N_x\times N_y$),
  $\boldsymbol{y}  \in \mathbb{C}^{N}$ denote the corresponding \textit k-t space measurement, where $N$ is the total number of points in the $k$-t space ($N= N_v\times N_c\times N_{TSL}$, $N_c$ denotes the coil number ). The forward model for $\text T_{1\rho}$ mapping is given by 
\begin{equation}
\begin{aligned}
    \boldsymbol{y} = E\boldsymbol{x} + \eta,
\end{aligned}
\end{equation}
where $\eta$ denotes the measurement noise and $E$ denotes the encoding operator given by $E=\mathcal{M}FC$~\cite{otazo2015low}, where $\mathcal M$ is the undersampling operator that undersamples $k$-space data for each $\text T_{1\rho}$-weighted image. $F$ denotes the Fourier transform operator. $S$ denotes an operator which multiplies the coil sensitivity map by each $\text T_{1\rho}$-weighted image coil-by-coil. The general formulation for recovering $\boldsymbol x$ from its undersampled measurements can be formulated as
\begin{equation}
\label{measure model}
\begin{aligned}
  \begin{gathered}
\underset{\boldsymbol{x} }{\operatorname{\arg \min}} \frac{1}{2}\left \|E \boldsymbol{x}-\boldsymbol{y} \right\|_{F}^{2} +\lambda R(\boldsymbol x) 
\end{gathered}, 
\end{aligned}
\end{equation}
where $\left\| \cdot \right\|_{F}$ denotes the Frobenius norm, $R(\boldsymbol x)$ denotes a combination of regularizations, and $\lambda$ is the regularization parameter.

\subsection{Self-consistency of $k$-t space prior}
Calibration-based PI approaches reconstruct images by enforcing the self-consistency of multi-coil $k$-space data. These approaches estimate a linear relationship between intra- and inter-coil $k$-space data from a tiny portion of fully sampled $k$-space center, namely ACS, and apply the relationship to reconstruct missing lines. The process can be expressed as an operator form:
\begin{equation}
\label{spirit function 1}
\begin{aligned}
\begin{gathered}
\text { minimize }\|(\mathcal{G}-\mathcal{I}) \hat{\boldsymbol y}\|_F^2 
\text { s.t. }\|\mathcal{M} \hat y-\boldsymbol y\|_F^2 \leq \varepsilon,
\end{gathered}
\end{aligned}
\end{equation}
where $\hat{\boldsymbol y}$ denotes the $k$-t space data for all coils. Unlike the SPIRiT method~\cite{lustig2010spirit}, here $\mathcal{G}$ is an operator convoluving the $k$-t space data with a series of calibration kernels that are estimated from the ACS including the temporal neighborhood of $k$-space sampling (shown in Fig.~\ref{fig1}(e) ). $\mathcal I$ is the identity operator, and $(\mathcal{G}-I)$ is often referred to as null-space kernel. 
Similar to Eq.~\ref{measure model}, additional penalty term can also be added to Eq.~\ref{spirit function 1} to further regularization the optimization problem~\cite{zhang2020image}. 

\subsection{Signal relaxation prior}
\begin{sloppypar}
In QMR, due to the exponentially decaying of signal relaxation property, the temporal evolution of image signals in the presence of chemical shifts, field inhomogeneity, and multiple relaxation mechanisms can be modeled as a linear combination of exponentials~\cite{peng2016accelerated,nguyen2012denoising,liu2023accelerating}. Let $S\left(\boldsymbol{v_s}, \textit {t}_{m}\right)$ denotes the temporal signal in $\boldsymbol{x}$, ${\boldsymbol{v_s}=[v_{x}^{h},v_{y}^{w}]_{h = 1,2,\cdots,N_x,w = 1,2,\cdots,N_y}}$ indicates the spatial coordinate, $\textit {t}_{m}$ is the \textit mth TSL, the signals from each voxel along the temporal direction can be used to form a Hankel matrix:
\begin{equation}
\label{Hankel matrix}
\begin{aligned}
\setlength{\arraycolsep}{1pt}
 \begin{array}{lc}
\boldsymbol H= \begin{bmatrix}
 S(\boldsymbol{v_s},t_1)& S(\boldsymbol{v_s},t_2)& \cdots& S(\boldsymbol{v_s},t_k) \\ 
 S(\boldsymbol{v_s},t_2)& S(\boldsymbol{v_s},t_3)& \cdots & S(\boldsymbol{v_s},t_{k+1}) \\ 
 \vdots & \vdots &\vdots& \vdots \\ 
 S(\boldsymbol{v_s},t_{N_{\textit{TSL}}-k+1})& S(\boldsymbol{v_s},t_{N_{\textit{TSL}}-k+2})& \cdots & S(\boldsymbol{v_s},t_{N_{\textit{TSL}}}) \end{bmatrix},
\end{array}
\end{aligned}
 \end{equation}
with a low rank property (i.e., $\text {rank}(\boldsymbol H) \ll N_\textit{TSL}$, where $\text{rank}(\boldsymbol H)$ denotes the rank of matrix $\boldsymbol H$), and the theoretical proof can be found in~\cite{peng2016accelerated}. The Hankel matrix is designed as square as possible, where $k$ is selected as the nearest integer greater than or equal to half the number of TSLs. The Hankel matrix construction process can be expressed as $\boldsymbol{H}_i = \mathcal{H}_i(\boldsymbol{x}), i = [1, 2, \cdots, N_v]$, where $\mathcal{H}$ is an operator that converts the temporal signals in $\boldsymbol{x}$ at fixed spatial locations to a Hankel matrix according to Eq.~\ref{Hankel matrix}.
\end{sloppypar}

\subsection{Overall framework of LINEAR}
\par Figure~\ref{fig1} shows the flow diagram of the proposed LINEAR method. In this study, the $\text T_{1\rho}$-weighted images $\boldsymbol{x}$ are modeled as a continuous function parameterized by a neural network $f_{\theta}$. This network is implemented as a coordinate-based MLP parameterized by weights $\theta$, with an $n_{e}$-dimensional input and a 2-dimensional output representing the real and imaginary components of the complex-valued $\text T_{1\rho}$-weighted images (Figure~\ref{fig1}(a)). Similar to~\cite{feng2023imjense}, the coordinates are normalized to the range of [-1, 1]. The forward model in Eq.~\ref{measure model} can be converted into optimizing the weights in $f_{\theta}$ using the following formula:
\begin{equation}
\label{INR model}
  \begin{gathered}
\theta^*=\underset{\theta }{\operatorname{\arg \min}} \frac{1}{2}\left \|E f_{\theta}(\gamma (\boldsymbol v))-\boldsymbol{y} \right\|_{F}^{2} +\lambda R(f_{\theta}(\gamma (\boldsymbol v))) 
\end{gathered}  
\end{equation}
To characterize the signal relaxation prior and the self-consistency of $k$-t space prior, two constraints are used to train the network $f_{\theta}$ by minimizing the following loss function:
\begin{equation}
\begin{aligned}
\theta^*=\arg\min_\theta \mathcal{L}_{\theta},\quad \mathcal{L}_{\theta} = \mathcal{L}_{DC} + \lambda_1\mathcal{L}_{HK} + \lambda_2\mathcal{L}_{SC}
\label{loss function 2}
\end{aligned}
\end{equation}
where $\mathcal{L}_{DC}$ is the data consistency term, $\mathcal{L}_{HK}$ is the low rank of Hankel 
matrices term, and $\mathcal{L}_{SC}$ is the self-
consistency of $k$-t space term. For the data consistency term, since the magnitudes of the low-frequency elements in $k$-space are several orders of magnitude greater than those of the high-frequency elements, we used the normalized $L_1$-norm loss instead of the $L_2$ norm loss for better high-frequency performance~\cite{yaman2020self}. For the self-consistency of the $k$-t space term, inspired by~\cite{santelli2014radial}, each point in the $k$-t space can be represented as a weighted sum of $k$-t points within a predefined full neighborhood across all coils and adjacent TSLs, which further exploits the temporal correlations of $k$-space. Consequently, the three terms in Eq.~\ref{loss function 2} are redesigned to characterize the signal relaxation and the self-consistency of $k$-t space data priors using the following formula:
\begin{equation}
\label{loss function 3}
\left\{
\begin{aligned}
   \mathcal{L}_{DC} &= \frac{\left\|Ef_\theta(\gamma (\boldsymbol v))-\boldsymbol y\right\|_1}{\left\|Ef_\theta(\gamma (\boldsymbol v))\right\|_1} \\
\mathcal{L}_{HK} &=\frac{1}{N_v}\sum_{i}^{N_v}\left\| \mathcal{H}_i(f_\theta(\gamma (\boldsymbol v))) \right\|_* \\
\mathcal{L}_{SC} &=\frac{1}{N}\left\| (\mathcal{G-I})Ef_\theta(\gamma (\boldsymbol v))\right\|_2^2
\end{aligned}
\right.
\end{equation}

\par Once the network is well-trained, it learns a representation of the $\text T_{1\rho}$-weighted image series, allowing the generation of predicted images by applying the trained network across all coordinates in the spatial and temporal fields. 
In this study, we introduced an additional step by replacing the predicted $k$-space at data sampling locations with the acquired $k$-space data. The final reconstructed images $\hat {\boldsymbol{x}}$ are then obtained by
\begin{equation}
\label{data consistency projection}
\begin{gathered}
\hat {\boldsymbol{x}}= E^*((\mathcal{I-M})Ef_{\theta^*}(\gamma (\boldsymbol v))+\boldsymbol{y})
\end{gathered}
\end{equation}
where $E^*$ denotes the adjoint operation of $E$. The whole training process is illustrated in Fig.~\ref{fig1} (a).

\section{Implementation}
\subsection{Data acquisition}
 \par The proposed method was evaluated on a fully sampled $\text {T}_{1\rho}$ mapping 
 dataset of the brain from six volunteers and a prospectively undersampled dataset collected from five volunteers. 
 All data were collected on a 3T scanner (TIM TRIO, Siemens, Erlangen, Germany) with a 12-channel head coil using the fast spin echo (FSE) sequence with a spin-lock frequency of 500 Hz. The local institutional review board approved the experiments, and informed consent was obtained from each volunteer. The imaging parameters were: FOV = 230 × 230 $\text {mm}^{2}$, imaging matrix = 384 × 384, slice thickness = 5 mm, TE/TR = 5.8 ms/4 s, echo train length = 16, TSLs = 1, 20, 40, 60, 80 ms. The average scan time for each fully sampled data was 5.4 mins, and 0.9 mins for the prospective undersampled data. 

In this study, all datasets were undersampled along the phase-encoding dimension using a pseudo-random undersampling acquisition scheme to evaluate the performance of reconstruction methods. Specifically, the phase-encoding lines in the center of the $k$-space were fully sampled, while the outer lines were undersampled using a golden-angle Cartesian sampling technique~\cite{rich2020cartesian}. Different sampling masks were used for each TSL. In the retrospective datasets, the fully sampled $k$-space data were undersampled using masks with net acceleration factors (R) of 6, 10, and 14. In the prospective study, each undersampled dataset was acquired using a pseudo-random undersampling pattern with R = 5.76. The coil sensitivity maps were calculated from the fully-sampled central $k$-space data using ESPIRiT~\cite{uecker2014espirit}.
 
\subsection{Network architecture and training }
Figure~\ref{fig1} (b) shows a schematic illustration of the network architecture used in this study. A 9-layer MLP based on the SIREN network~\cite{sitzmann2020implicit} was utilized for learning INR. All layers, except for the last one, utilized the periodic sine activation function. The network is designed to accommodate an embedded matrix and skip connections~\cite{huang2017densely}. Its input consists of a 3-dimensional coordinate grid, initialized within the range \([-1, 1]\). This grid is mapped to a higher-dimensional space using the Fourier feature mapping, resulting in 256 features, which serve directly as the input to the MLP. Therefore, the first layer of the network receives 256 input channels. To enhance the network's fitting capabilities, skip connections are incorporated into the intermediate layers (3rd, 5th, and 7th). In these layers, the embedded matrix is concatenated with the output from the previous layer, resulting in 512 input channels for the next layer. The final layer of the network outputs 2 channels, representing the real and imaginary parts of the predicted images.

\begin{sloppypar}
To accelerate the convergence of the network, we first conducted a pre-training process using an undersampled measurement with R = 6 from a randomly selected volunteer. Initially, the network's weights were set using values sampled from uniform distributions. Specifically, the first layer’s weights were initialized with values from~\( U(-1/n_e, 1/n_e)\), where~\(n_e = 256\), while subsequent layers’ weights were initialized with values from \( U(-\omega_0 \sqrt{6/n_l}, \omega_0 \sqrt{6/n_l}) \), where \(\omega_0 = 30\) is a frequency factor used to enhance the fitting ability of the sine network, and \( n_l \) is the number of input channels for each layer. The initial learning rate for this process was set to 0.0001. The network was trained until full convergence to obtain a well-optimized set of parameters. Then these parameters in the pre-trained network were used as an initialization in the network training for other subsequent reconstruction experiments. The initial learning rate was set to 0.00035 and decayed by 50\% every 700 iterations. The training process included 3500 iterations, with rapid convergence typically occurring after 500 iterations. This approach significantly reduced the overall convergence time. For different acceleration factors (R = 6, R = 10, R = 14), the regularization parameter \(\mathbf{\lambda_{1}}\) and \(\mathbf{\lambda_{2}}\) were set to (15.8, 10.7, 13.8) and (1277, 1480, 1538), respectively, determined using the Ray Tune~\cite{Liaw2018TuneAR}, an optimization framework that integrates Optuna.
\end{sloppypar}

All experiments were conducted on an Ubuntu 20.04 operating system equipped with an NVIDIA A100 Tensor Core (GPU, 80 GB memory) in the open PyTorch 1.10 framework~\cite{paszke2019pytorch} with CUDA 11.8 and CUDNN support. For the proposed LINEAR, ADAM optimizer~\cite{kingma2014adam} with ${\beta}_1 = 0.9$, ${\beta}_2 = 0.999$ was chosen for optimizing the loss function in Eq.~\ref{loss function 2}.

\subsection{\texorpdfstring{$\text{T}_{1\rho}$ map estimation}{T1rho map estimation}}
To estimate the $\text {T}_{1\rho}$ map, a signal relaxation model was used by fitting the reconstructed $\rm T_{1\rho}$-weighted images with different TSLs pixel-by-pixel:
\begin{equation}
\label{momo_expo}
M_{k}=M_{0} \exp \left(-{TSL}_k/ T_{1 \rho}\right)_{k=1,2, \ldots, N_{\textit {TSL }}},
\end{equation}
where $M_0$ denotes the baseline image intensity without applying a spin-lock pulse, and $M_k$ is the signal intensity for the \textit{k}th TSL image. $\text T_{1\rho}$ map was estimated using the nonlinear least-squares fitting method with the Adam optimizer and an iteration number of 50000.
\subsection{Comparison methods}
For the purpose of performance comparison, the proposed LINEAR method and five state-of-the-art methods were used to reconstruct the undersampled $k$-space data. These methods inlude the subject-specific implicit representation for joint coil sensitivity and image estimation (IMJENSE) method which utilizes the sparsity prior of images~\cite{feng2023imjense}, the zero-shot self-supervised learning (ZS-SSL) method~\cite{yaman2021zero} which performs subject-specific training of physic-guided deep learning reconstruction in an unrolled way, the ConvDecoder method~\cite{darestani2021accelerated} which utilizes an unsupervised neural netork for accelerated MRI, the ESPIRiT method with a $L_1$ wavelet regularization which enforces the self-consistency of $k$-space data~\cite{lustig2010spirit}, and the model-driven low rank and sparsity priors method (MORASA)~\cite {peng2016accelerated} which enforces the low rank of the Hankel matrix over all spatial locations and sparsity of the image matrix. The quantitative assessment of reconstructed $\text T_{1\rho}$-weighted images used the peak signal to noise ratio (PSNR), structural similarity (SSIM) index, and normalized root mean square error (NRMSE).

\begin{figure*}[!ht]
\centering \includegraphics[width=2\columnwidth]{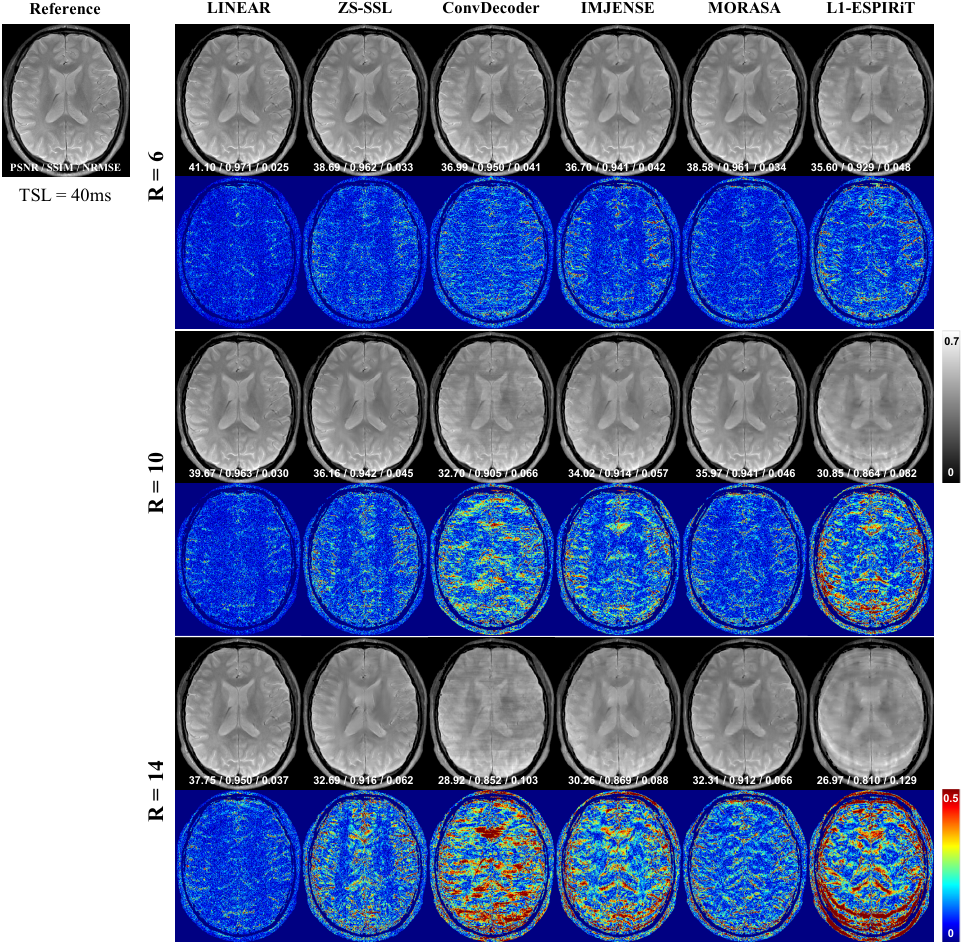}
\caption{ Reconstructed $\text T_{1\rho}$-weighted images from one retrospective dataset at TSL = 40 ms with acceleration factors R = 6, 10 and 14 using the LINEAR, ZS-SSL, ConvDecoder, IMJENSE, MORASA, and L1-ESPIRiT methods. The corresponding error images for the reference image reconstructed from the fully sampled data are also shown. The error images are amplified by eight for visualization. The quantitative metrics including PSNR, SSIM, and NRMSE are shown at the bottom of each reconstructed image.}
\label{fig2}
\end{figure*}
\subsection{Ablation studies}
We conducted ablation experiments to analyze the effects of the three configurations in Eq.~\ref{loss function 2} on the reconstruction. Three loss functions were designed to train the network and the following models were used to reconstruct the image respectively:\\
\underline{LINEAR-DC:}
\begin{equation}
\theta^*=\arg\min_\theta \mathcal{L}_{DC}
\label{loss function 4}
\end{equation}
\underline{LINEAR-HK:}
\begin{equation}
\theta^*=\arg\min_\theta \mathcal{L}_{DC} + \lambda_1\mathcal{L}_{HK}
\label{loss function 5}
\end{equation}
\underline{LINEAR-SC:}
\begin{equation}
\theta^*=\arg\min_\theta \mathcal{L}_{DC} + \lambda_2\mathcal{L}_{SC}
\label{loss function 6}
\end{equation}
The ablation experiments were implemented on the retrospective datasets, with the acceleration factors of R = 6, R = 10, and R = 14.

\begin{figure*}[!ht]
\centering \includegraphics[width=2\columnwidth]{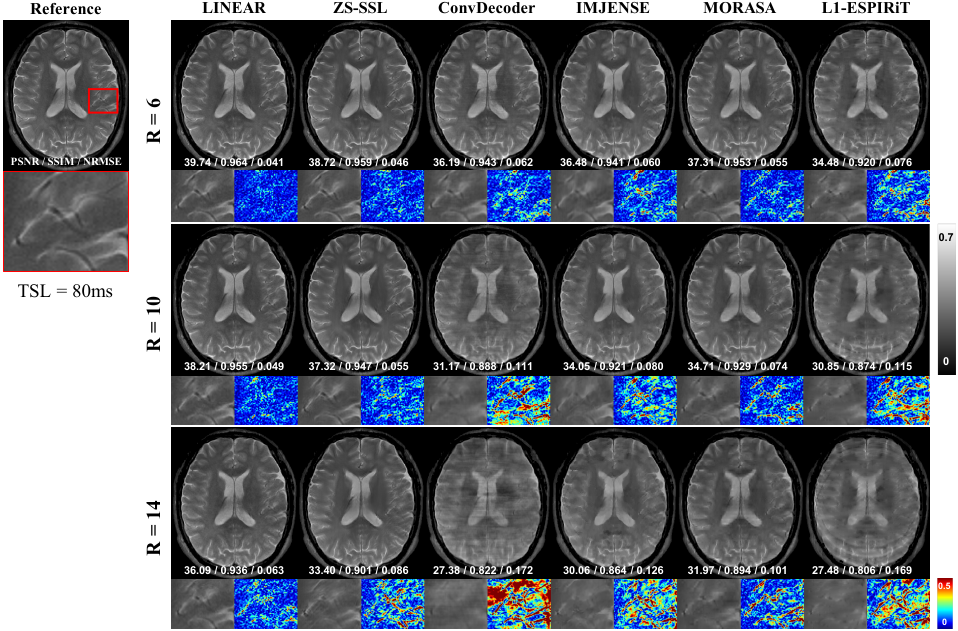}
\caption{Reconstructed $\text T_{1\rho}$-weighted images from one retropective dataset at TSL = 80 ms with acceleration factors R = 6, 10, and 14 using the LINEAR, ZS-SSL, ConvDecoder, IMJENSE, MORASA, and L1-ESPIRiT methods. The zoom-in figures of region of interest (denoted by the red box) are shown at the bottom of each reconstructed image. The corresponding error images for the reference image reconstructed from the fully sampled data are also shown. The error images are amplified by eight for visualization. The quantitative metrics including PSNR, SSIM, and NRMSE are shown at the bottom of each reconstructed image.}
\label{fig3}
\end{figure*}

\begin{table*}[!hbt]
\caption{Comparisons of different methods on the retrospective data with various acceleration factors (AF). The best results are shown in bold.}
\centering \renewcommand{\arraystretch}{1}
\begin{tabular}{cc|c c c c c c}
\Xhline{3\arrayrulewidth} \hline \hline
\begin{tabular}{@{}c@{}} AF \end{tabular} & Metrics & LINEAR & ZS-SSL & ConvDecoder & IMJENSE & MORASA & L1-ESPIRiT \\
\Xhline{3\arrayrulewidth}
\multirow{3}{*}{\centering{R = 6}} & PSNR & \textbf{40.28$\pm$0.52} & 38.31$\pm$0.51 & 36.30$\pm$0.47 & 36.44$\pm$0.44 & 37.73$\pm$0.59 & 35.05$\pm$0.66 \\
& SSIM & \textbf{0.966$\pm$0.004} & 0.959$\pm$0.006 & 0.945$\pm$0.003 & 0.938$\pm$0.007 & 0.956$\pm$0.004 & 0.923$\pm$0.010 \\
& NRMSE & \textbf{0.028$\pm$0.007} & 0.035$\pm$0.007 & 0.045$\pm$0.010 & 0.044$\pm$0.011 & 0.038$\pm$0.009 & 0.052$\pm$0.015 \\
\Xhline{3\arrayrulewidth}

\multirow{3}{*}{\centering{R = 10}} & PSNR & \textbf{38.71$\pm$0.62} & 36.14$\pm$0.68 & 32.13$\pm$0.53 & 33.88$\pm$0.22 & 35.12$\pm$0.59 & 30.93$\pm$0.38 \\
& SSIM & \textbf{0.956$\pm$0.006} & 0.937$\pm$0.010 & 0.898$\pm$0.005 & 0.915$\pm$0.006 & 0.934$\pm$0.004 & 0.869$\pm$0.011 \\
& NRMSE & \textbf{0.034$\pm$0.009} & 0.045$\pm$0.008 & 0.073$\pm$0.022 & 0.059$\pm$0.014 & 0.051$\pm$0.013 & 0.083$\pm$0.021 \\
\Xhline{3\arrayrulewidth}

\multirow{3}{*}{\centering{R = 14}} & PSNR & \textbf{36.86$\pm$0.67} & 33.22$\pm$0.62 & 28.36$\pm$0.52 & 30.54$\pm$0.48 & 31.99$\pm$0.74 & 27.83$\pm$0.95 \\
& SSIM & \textbf{0.940$\pm$0.008} & 0.913$\pm$0.014 & 0.840$\pm$0.010 & 0.881$\pm$0.017 & 0.901$\pm$0.007 & 0.824$\pm$0.025 \\
& NRMSE & \textbf{0.042$\pm$0.012} & 0.064$\pm$0.014 & 0.113$\pm$0.033 & 0.087$\pm$0.023 & 0.073$\pm$0.015 & 0.119$\pm$0.032 \\
\Xhline{3\arrayrulewidth}
\end{tabular}
\label{table of quantitative results}
\end{table*}

\begin{figure*}[!htb]
\centering \includegraphics[width=2\columnwidth]{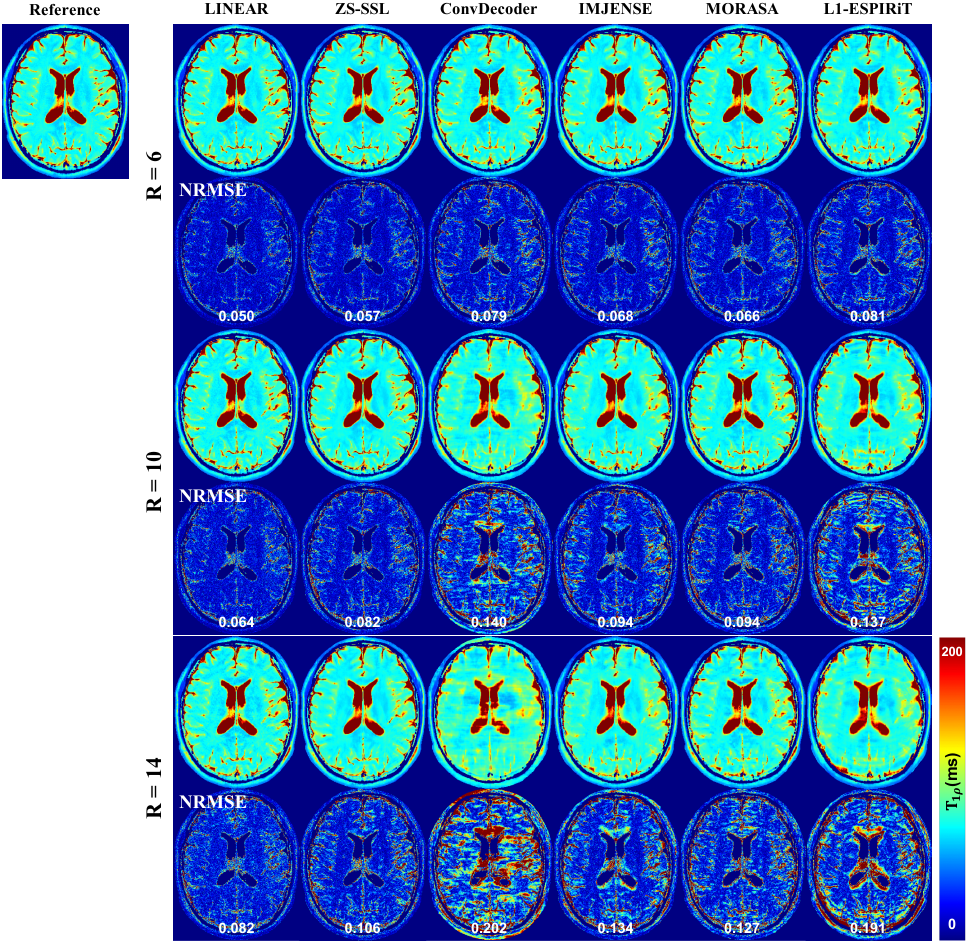}
\caption{Reconstructed $\text T_{1\rho}$ maps estimated from the same retrospective dataset in Fig.~\ref{fig2} and Fig.~\ref{fig3} with acceleration factors R = 6, 10, and 14 using the LINEAR, ZS-SSL, ConvDecoder, IMJENSE, MORASA, and L1-ESPIRiT methods. The corresponding error images for the reference estimated from fully sampled data are also shown. The NRMSEs are shown at the bottom of the error maps.}
\label{fig4}
\end{figure*}

\section{Experimental Results}
\subsection{Retrospective reconstruction}
The $\text T_{1\rho}$-weighted images at TSL = 40 ms from one volunteer reconstructed using the LINEAR, ZS-SSL, ConvDecoder, IMJENSE, MORASA, and L1-ESPIRiT methods with different acceleration factors of R = 6, 10, and 14, respectively, are shown in Fig.~\ref{fig2}. The error images between the reconstructed and reference images are displayed under the reconstructions with PSNR, SSIM, and NRMSE placed below the reconstructed images. 

In Fig.~\ref{fig2}, all methods, excluding L1-ESPIRiT, suppress aliasing artifacts well and offer acceptable results at a low acceleration factor of R = 6. The quality of reconstructions using traditional CS and PI methods degrades significantly, suffering from blurring and aliasing artifacts with increased acceleration factors. The images reconstructed using the INR-based IMJENSE method show a significant loss of image details and considerable blurring at high acceleration factors R = 10 and 14. Noticeable artifacts remain in the reconstructions by the unsupervised ConvDecoder method, which lacks prior knowledge during training. The unrolling-based ZS-SSL method, with its learnable regularization prior, produces visually better reconstructed images than ConvDecoder. Nevertheless, this prior learned from the zero-shot data may not be optimal, limiting the reconstruction performance. The proposed LINEAR method shows few reconstruction errors even at R = 14. It preserves more fine details by integrating both prior knowledge of the signal relaxation and the self-consistency of $k$-t space data into the network. LINEAR also has the best quantitative performance compared to the other five methods for all acceleration factors.

For better visualization and demonstrating the superiority of the LINEAR method, Fig.~\ref{fig3} shows the $\text T_{1\rho}$-weighted images at TSL = 80 ms and the zoom-in images of region of interest (ROI) from the same volunteer using the above six methods at different acceleration factors. The corresponding error images in the ROI are also shown. We can observe that the reconstructed images using the proposed LINEAR method show highly similar anatomical details as the reference image. As pointed in the zoom-in images, certain image details with intricate structures, such as blood vessels, are well preserved in the LINEAR reconstructions, even at a high acceleration factor of up to 14.

\begin{figure*}[!ht]
\centering \includegraphics[width=1.7\columnwidth]{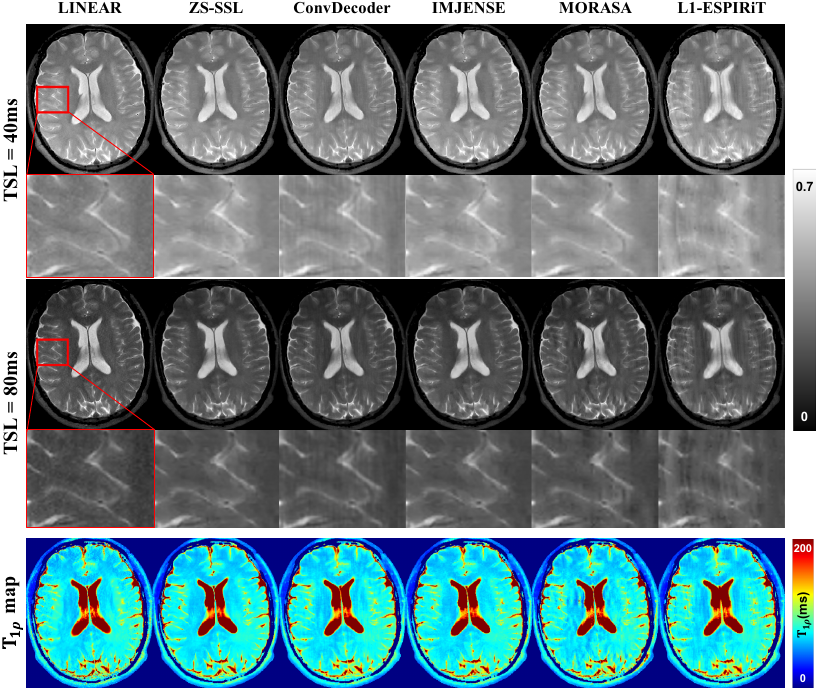}
\caption{Reconstructed $\text T_{1\rho}$-weighted images from one prospective dataset at TSL = 40 ms and TSL = 80 ms with R = 5.76 using the LINEAR, ZS-SSL, ConvDecoder, IMJENSE, MORASA, and L1-ESPIRiT methods. The zoom-in figures of region of interest (denoted by the red box) are shown at the bottom of each reconstructed image. The estimated $\text T_{1\rho}$ maps from the reconstructed $\text T_{1\rho}$-weighted images are also shown.}
\label{fig5}
\end{figure*}

\begin{table*}[!hbt]
\caption{Ablation study of the effects of the signal relaxation and the self-consistency of $k$-t space priors on reconstruction with different acceleration factors (AF). The best results are shown in bold.}
\centering \renewcommand{\arraystretch}{1} 
\begin{tabular}{cc|c c c c}
\Xhline{3\arrayrulewidth} \hline \hline
\begin{tabular}{@{}c@{}} AF \end{tabular} & Metrics & LINEAR-DC & LINEAR-SC & LINEAR-HK & LINEAR \\

\Xhline{3\arrayrulewidth}
\multirow{3}{*}{\centering{R = 6}} & PSNR & 37.03$\pm$0.57 & 37.34$\pm$0.62 & 40.14$\pm$0.52 & \textbf{40.28$\pm$0.52} \\
& SSIM & 0.938$\pm$0.007 & 0.939$\pm$0.007 & 0.965$\pm$0.004 & \textbf{0.966$\pm$0.004} \\
& NRMSE & 0.042$\pm$0.011 & 0.040$\pm$0.011 & 0.029$\pm$0.007 & \textbf{0.028$\pm$0.007} \\
\Xhline{3\arrayrulewidth}

\multirow{3}{*}{\centering{R = 10}} & PSNR & 32.99$\pm$0.26 & 33.31$\pm$0.24 & 37.78$\pm$0.73 & \textbf{38.71$\pm$0.62} \\
& SSIM & 0.904$\pm$0.007 & 0.906$\pm$0.007 & 0.955$\pm$0.005 & \textbf{0.956$\pm$0.006} \\
& NRMSE & 0.065$\pm$0.014 & 0.063$\pm$0.014 & 0.037$\pm$0.008 & \textbf{0.034$\pm$0.009} \\
\Xhline{3\arrayrulewidth}

\multirow{3}{*}{\centering{R = 14}} & PSNR & 29.32$\pm$0.38 & 30.15$\pm$0.55 & 36.45$\pm$0.70 & \textbf{36.86$\pm$0.67} \\
& SSIM & 0.862$\pm$0.023 & 0.864$\pm$0.022 & 0.935$\pm$0.008 & \textbf{0.940$\pm$0.008} \\
& NRMSE & 0.100$\pm$0.025 & 0.092$\pm$0.027 & 0.044$\pm$0.011 & \textbf{0.042$\pm$0.012} \\
\Xhline{3\arrayrulewidth}
\end{tabular}
\label{table of ablation}
\end{table*}

Table~\ref{table of quantitative results} shows the mean and standard deviation values of the quantitative etrics for reconstructed images. With the increase of accelerate factors, the PSNR and SSIM decline and NRMSE growth are observed in all methods. The LINEAR method achieves the best performance among the six methods, with the highest PSNR and SSIM values and the lowest NRMSE values at all acceleration factors. The evaluation criteria are consistent with visual inspections mentioned above. The estimated $\text T_{1\rho}$ maps from the reconstructed $\text T_{1\rho}$-weighted images and error maps corresponding to the reference $\text T_{1\rho}$ map are shown in Fig.~\ref{fig4}. The NRMSE values are placed below the error maps. The proposed LINEAR method shows comparable $\text T_{1\rho}$ maps to the reference with the smallest NRMSE compared to the other five methods, and the NRMSEs are less than 9\%, which further confirm the superiority of the LINEAR method.

\subsection{Prospective reconstruction}
Figure~\ref{fig5} shows the prospectively reconstructed $\text T_{1\rho}$-weighted images at TSL = 40, 80 ms, and the $\text T_{1\rho}$ maps from another volunteer with R = 5.76 using the LINEAR, ZS-SSL, ConvDecoder, IMJENSE, MORASA, and L1-ESPIRiT methods. Noticeable aliasing artifacts remain in reconstructions by L1-ESPIRiT and ConvDecoder. The ZS-SSL, IMJENSE and MORASA method produce blurring. By contrast, LINEAR retains promising image resolution and fine details. For better visualization of their difference, we offer zoom-in images of reconstructions, where one can observe relatively stronger artifacts in the L1-ESPIRiT and ConvDecoder images, and somewhat blurring in ZS-SSL, IMJENSE and MORASA images. The proposed LINEAR show a great capability to suppress artifacts, preserve fine details, and present reliable reconstructions.

\subsection{Results of Abalation study}
Table~\ref{table of ablation} shows the results of quantitative metrics for reconstructed images at all TSLs using the loss functions in Eq.~\ref{loss function 4},~\ref{loss function 5} and~\ref{loss function 6} respectively with R = 6, R = 10, and R = 14. The proposed LINEAR method achieves the highest PSNR, SSIM and the lowest NRMSE in all cases,
implying better reconstructed images when characterizing the signal relaxation prior and the self-consistency of $k$-t space prior jointly.

\section{Discussion}
In this study, we proposed a method which learns a implicit neural representation of the MR images driven by the signal relaxation prior, and the self-consistency of $k$-t space prior. The proposed LINEAR method was applied to accelerated MR $\text T_{1\rho}$ mapping to enable undersampling factors that go beyond the limit of traditional PI and CS reconstructions with the acceleration factor up to 14, while suppressing the aliasing artifacts and preserve image details. Comparative experiments on retrospective and prospective datasets were carried out to verify the effectiveness of the proposed method. The LINEAR method significantly outperforms other unsupervised methods and traditional PI and CS-based methods, with image reconstruction comparable to the fully sampled reference. 
\subsection{Loss function}
Many constraints in the image domain or $k$-space domain, such as TV on the image, low rank of the Casorati matrix of the $\text T_{1\rho}$-weighted image series, or low rank of block Hankel matrices in $k$-space, can be utilized in the loss function to train the network. In our initial attempts, using these constraints was challenging to remove aliasing artifacts and often led to blurring in the reconstructions, particularly when the acceleration factor was high (i.e., R = 10). Therefore, we leveraged the physical priors in our final implementation, including the signal relaxation prior, enforced by constructing pixel-wise low rank Hankel matrices along the TSL dimension, and the self-consistency in $k$-t space prior, enforced by constructing the consistency operator $\mathcal{G}$ using a convolution kernel estimated from $k$-t space data. These constraints can significantly improve the reconstruction performance by suppressing aliasing artifacts and preserving fine image details. More complex constraints, such as group low rank prior~\cite{bustin2019high}, may further improve the results but can also lead to higher computation complexity and longer running time. Considering the trade-off between the training time and reconstruction performance, we employed both the $\mathcal L_{SC}$ and $\mathcal L_{HK}$ terms in the loss function for this study.

\subsection{Parameter setting}
In the proposed LINEAR method, we utilized the SIREN network as the MLP, which can improve the fitting capacity for high-frequency information, as has been pointed in previous studies. The parameter $w_0$ is related to the high-frequency information of the reconstructed image. In addition, the two regularization parameters $\lambda_1$ and $\lambda_2$ in the loss function, also contribute to the reconstruction performance and need to fine tuned to achieve satisfactory reconstruction. We utilized the Ray Tune toolbox~\cite{Liaw2018TuneAR} to optimize the three hyperparameters. From numerous combinations, we selected the parameter settings that maximized the PSNR of the reconstructed images.
 
\subsection{Limitations and future work}
There are several limitations in this study. On one hand, the training time for LINEAR is relative long. The average time for each dataset is 16 minutes, which is longer than the IMJENSE method (14 mins), MORASA (10 mins), and L1-ESPRiT (20 secs), and shorter than the ZS-SSL (215 mins) and ConvDecoder (18 mins). The long training time significantly hinders the proposed method’s clinical application. One potential solution to this problem is to use the computational efficiency boosting methods, such as the hash encoding~\cite{muller2022instant}, or using multiple tiny MLPs~\cite{reiser2021kilonerf} to speeding up the optimization process. On the other hand, one Fourier feature embedding was employed to all the coordinates, including the spatial and temporal coordinates. Embedding the spatial and temporal coordinates separately~\cite{catalan2023unsupervised,kunz2023implicit}, or may lead to improved reconstruction, which will be explored in our future studies.

\section{Conclusion}
This study investigates the application of implicit neural representation in MR $\text T_{1\rho}$ mapping to enhance reconstruction outcomes. We propose a novel INR-based method, called LINEAR, that effectively exploits correlations among the multi-coil $k$-space data and among $\text T_{1\rho}$-weighted images arcoss the TSLs. Experimental results in both retrospective and prospective imaging scenarios show that the LINEAR method can improve the reconstruction results qualitatively and quantitatively, and achieves superior performance in artifacts suppression and image detail preservation than the state-of-the-art methods.

\section*{Acknowledgments}
\sloppy This study was supported in part by the National Key R\&D Program of China nos.2023YFA1011403, 2021YFF0501402, 2020YFA0712200, National Natural Science Foundation of China under grant nos.62322119, 62201561, 62206273, 12226008, 62125111, 62106252, 81971611, and 81901736, the Guangdong Basic and Applied Basic Research Foundation under grant no.2021A1515110540, and the Shenzhen Science and Technology Program under grant no.RCYX20210609104444089.

\bibliographystyle{IEEEtran}
\bibliography{main}


\end{document}